\documentclass[osajnl,,twocolumn,showpacs,superscriptaddress,10pt]{revtex4-1} %% use 11pt for Applied Optics

\usepackage{amsmath,amssymb,graphicx}
\usepackage{epstopdf}
\usepackage{siunitx}

\begin{document}

\title{$8
\times10^{-17}$ fractional laser frequency instability with a long room-temperature cavity}

\author{Sebastian H\"afner}
\author{Stephan Falke}
\thanks{Current address: TOPTICA Photonics AG, Lochhamer Schlag 19, 82166 Gr\"afelfing}
\author{Christian Grebing}
\author{Stefan Vogt}
\author{Thomas Legero}
\affiliation{Physikalisch-Technische Bundesanstalt (PTB), Bundesallee 100, 38116 Braunschweig, Germany}

\author{Mikko Merimaa}
%\affiliation{Centre for Metrology and Accreditation (MIKES), P.O. Box 9, FI-02151 Espoo, Finland}
\affiliation{VTT Technical Research Centre of Finland Ltd Centre for Metrology MIKES, P.O. Box 1000, FI-02044 VTT, Finland}

\author{Christian Lisdat}
\author{Uwe Sterr}
\affiliation{Physikalisch-Technische Bundesanstalt (PTB), Bundesallee 100, 38116 Braunschweig, Germany}

\begin{abstract}
We present a laser system based on a $\SI{48}{\centi\metre}$ long optical glass resonator. 
The large size requires a sophisticated thermal control and optimized mounting design. 
A self balancing mounting was essential to reliably reach sensitivities to acceleration of below $\Delta \nu / \nu < 2 \times 10^{-10}~/\mathrm{g}$ in all directions.
Furthermore, fiber noise cancellations from a common reference point near the laser diode to the cavity mirror and to additional user points (Sr clock and frequency comb) are implemented. 
Through comparison to other cavity-stabilized lasers and to a strontium lattice clock an instability of below $1 \times 10^{-16}$ at averaging times from $\SI{1}{\second}$ to $\SI{1000}{\second}$ is revealed.
\end{abstract}

% REPLACE WITH CORRECT OCIS CODES FOR YOUR ARTICLE
% NOTE: \ocis{} IS ALIASED TO \pacs{} BUT MUST
\maketitle %% required
% FORMAT THE TERMS CORRECTLY FOR EACH JOURNAL
%\section{Introduction}
Lasers with high frequency stability are urgently needed in various fields of physics and technology, e.g. for precision measurements, for operation of optical clocks \cite{ros08,ush15} or generation of low phase noise microwave signals \cite{bar05}. 
%Laser frequency noise can contribute to the noise floor in laser interferometers with unequal path length, and in optical clocks it limits the maximum interrogation time and thus the achievable resolution.
%blo14, 
The stability of today's best optical clocks \cite{hin13} is limited by laser frequency noise due to the Dick effect \cite{dic87}.
While alternative concepts for laser frequency stabilization like spectral hole-burning \cite{tho11a}, or active atomic clocks \cite{mei09} are currently under investigation, 
most ultrastable lasers are based on well isolated resonators.
%, which transforms length to frequency fluctuations. 
The best present room-temperature resonators provide fractional laser instabilities of $1$\textendash$2\times 10^{-16}$ \cite{nic12,jia11}, limited by the thermal Brownian noise from their constituents \cite{num04}, with the biggest contributions from mirror coatings.
This noise can be reduced by operation at cryogenic temperatures \cite{kes12a}, increasing the mode size \cite{ama13}, or using a longer cavity \cite{jia11}. 

Here we employ the latter approach and report on the design and evaluation of a $\SI{48}{\centi\metre}$ long room temperature resonator with a spacer made from ultra-low expansion glass (ULE). 
Compared to a cryogenic cavity, we can operate the system at ambient temperatures directly at the Sr lattice clock $\SI{698}{\nano\meter}$ transition \cite{fal14}, which simplifies the system and its operation.  
However, the use of long cylindrical and heavier cavities makes it much more difficult to 
reach the required suppression of optical length changes through forces induced by seismic vibrations and through thermal expansion caused by temperature fluctuations. 
In the following, we describe the vibration insensitive mounting, temperature control, optical setup and fiber noise cancellation system.

\begin{figure}[htbp]
\centerline{\includegraphics[width=0.95\columnwidth]{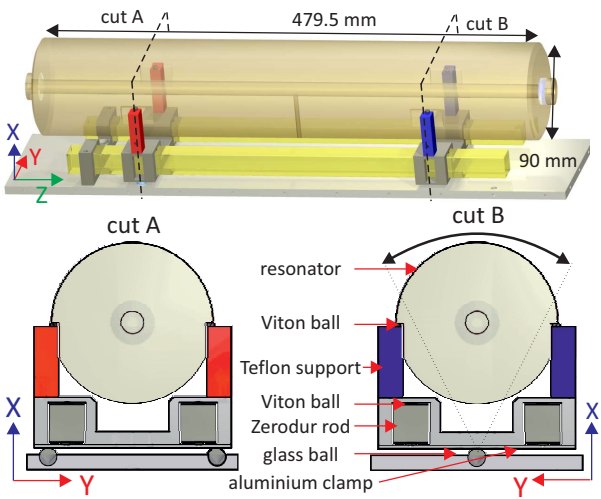}}
\caption{Resonator mounting design. The lower figures show cuts through the fixed (A) and rotatable (B) support.}
\label{fig:cavity design}
\end{figure}

We use a cylindrical ULE-glass spacer of $\SI{48}{\centi\metre}$ length and $\SI{9}{\centi\metre}$ diameter.
With two optically contacted fused silica mirrors (plane and radius of curvature $R=\SI{1}{\metre}$) we expect a thermal noise level \cite{kes12} of $5.4\times10^{-17}$. The system provides a cavity finesse $F = 282\,000$ ($\SI{1.1}{\kilo\hertz}$ linewidth) as measured from optical ring down. 
Before the final optical contact, the finesse was optimized by moving the mirrors to minimize the influence of coating defects.
ULE rings are attached to the rear sides of both mirrors to avoid the effect of the differential thermal expansion between the fused silica substrate and the ULE glass spacer \cite{leg10}. 
The resonator is held horizontally on four points (Fig.~\ref{fig:cavity design}) in small cutouts (length 40~mm, depth 4~mm) machined to the sides of the spacer \cite{naz06,web07,mil09}.
Accelerations acting on the resonator due to seismic noise both change the cavity length $L_\mathrm{geo}$ between the nominal centers of the mirrors (geometrical axis) and lead to tilts of the mirrors through bending of the spacer.
Mirror tilt leads to a change in the optical length if there is a mismatch between the geometrical axis and the optical axis.
The high symmetry of the resonator and mount largely suppresses the sensitivity of $L_\mathrm{geo}$ to accelerations in $y$ and $z$ directions.
In addition positions and dimensions of the cutouts were optimized by Finite Element Modeling (FEM) to minimize bending and the sensitivity along $x$.
%Because of symmetry of the resonator, accelerations perpendicular to the optical axis ($x$ or $y$ directions) do not change the length along the symmetry axis $L_\mathrm{sym}$ as long as the corresponding reaction forces at the mounting are also symmetric. 
%Mirror tilt through bending only leads to a change in the optical length in combination with a mismatch between the geometrical axis and its optical axis. 
%
%In vertical $(x)$ direction the symmetry is broken by the cutouts. Thus vertical acceleration can introduce a length change in $L_\mathrm{sym}$ through Poisson's effect.
%The positions and dimensions of the cutouts were optimized by Finite Element Modeling (FEM) to minimize both this effect and bending; it turned out that the optimum mounting position in the vertical (x) direction is still in the axial (yz) symmetry plane. 
%
%The mounting positions are also symmetric along the optical axis (z direction). Thus, if the four reaction forces are equal, the length $L_\mathrm{sym}$ is also not affected by accelerations along this direction.    
%
%The first effect directly translates to a change of the length $L_\mathrm{sym}$ while the bending in combination with a mismatch between the geometrical symmetry axis and the optical axis given trough the mirror position leads to a change of the optical eigenfrequency.
%Additionally Poisson's ratio effects caused through a geometrical offset between the vertical mounting position and the geometrical symmetry plane (yz) and e.g. vertical accelerations have an influence on the optical length, too. 
%
For a long and heavy spacer ($m=\SI{6.8}{\kilogram}$) as used here, the symmetry of the reaction forces along the optical axis ($z$) as well in the transverse horizontal direction ($y$) becomes more important, as already a force difference of \SI{100}{\nano\newton} between the mounting points (along $z$) leads to a relative length change of $1.3\times10^{-16}$. 
Assuming a typical vibration noise of $1~\mu$g, these reaction forces need to be balanced on a level of $10^{-3}$ to reach a fractional instability below $10^{-16}$. 
%For this symmetry, exactly the same elasticity at all four mounting positions has to be achieved. 
The elasticity at all four mounting positions has to be matched very accurately in order to obtain this level of balance of forces.
 
In our design the cavity is sitting on four Viton balls (diameter $\SI{4}{\milli\metre}$), which are placed on top of four Teflon posts. Through height imperfections of the posts, the Viton pieces get compressed differently and the shearing elasticity differs. Thus, the requirements given above are violated. 
%of the Viton balls in both horizontal directions. accelerations introduce different shearing forces acting different on each Viton ball introduced by . 
%As Viton pieces get compressed nonlinearly with the cavity weight \cite{tat91, liu98} also their susceptibility along the other directions changes accordingly.
%Hight 
%Small changes due to mechanical tolerances cannot be avoided completely, which lead to differential shearing forces in the Viton balls introduced through horizontal accelerations. 
%
To remove the height imperfections and realize an effective three point mount, a self balancing mount is employed (Fig. \ref{fig:cavity design}). 
Near one end of the resonator (cut B), the two Teflon posts supporting the resonator are mounted on an aluminum clamp on top of a single glass sphere, allowing the structure to rotate around the $z$-axis.
One the other end (cut A) rotation of the structure around the $z$-axis is prevented by two glass spheres underneath the clamp.
Thus, the four posts is automatically leveled and the forces at the support points are equalized by this movement of the whole cavity.
%Thus the top of the Teflon posts can move vertically, which realizes an effective three point mount where the forces get equally distributed by a corresponding %movement of the whole cavity. 

To isolate the mounting structure from the thermal expansion of the aluminum base plate, two Zerodur glass rods are inserted between the clamp and the base plate (Fig.~\ref{fig:cavity design}). 
Viton balls between the clamps and the glass rods ensure a strong friction force to prevent a horizontal movement of the clamps while providing the necessary flexibility for the balanced mount. 
By experimentally varying the distance between the Viton balls that hold the resonator, 
%As in practice, there is always some small offset between optical and geometrical axis, this way the length change from remaining axial expansion and from bending can be largely canceled.
%
%highly reproducible values could be obtained that after the optimization of the Viton ball displacement, a vibration sensitivities of 
the vibration sensitivities $\kappa_i$ were optimized to
$\kappa_z = 1.7\times 10^{-10}~\mathrm{/g}$, 
$\kappa_y = 0.5\times 10^{-10}~\mathrm{/g}$ and
$\kappa_x = 1.5\times 10^{-10}~\mathrm{/g}$; all measured at $0.7~$Hz,   
in contrast to poorly reproducible values around $4\times 10^{-8}~/\rm{g}$ which were initially obtained without balanced mount. 
%\section{Thermal design}

We measured the thermal expansion coefficient $\alpha$ of the resonator as
%\begin{equation}
$
	\alpha(T) = 2\times 10^{-9}~\mathrm{K}^{-2} (T-T_0)
$
%\end{equation}
with a zero-crossing temperature of $T_0=-0.24\,^\circ \mathrm{C}$.
Thus, if operated within $50$~mK of $T_0$, a temperature stability of better than $1~\mu$K is required to reach a fractional frequency instability of below $10^{-16}$. The large dimensions of the setup can easily lead to large temperature gradients that fluctuate with the ambient temperature, which exacerbates the control of the average temperature of the resonator at the required level.
To achieve the required stability we start by stabilizing the temperature of the vacuum chamber by  thermoelectric coolers (TEC) close to ambient temperature. 
%The elements are mounted between the bottom of the chamber that was machined from a block of aluminum and the aluminum base plate of the passive vibration isolation table, which serves as a heat sink. 

Inside the vacuum chamber, the temperature of the first heat shield is stabilized at $T_0$ using TECs mounted between the shield and the chamber.
To reduce the influence of thermal gradients on the cavity, this first shield is temperature stabilized in three sections along the axial direction ($z$). 
For each section a weighted temperature from thermistors at the top and bottom ($\approx1:5$) of the shield is used as input to a temperature control loop. 
The relative contributions of the sensors were determined experimentally by introducing gradients in the vacuum chamber and minimizing the remaining temperature fluctuation of the second shield.
Further inside, a third heat shield surrounds the resonator. 
The polished aluminum heat shields act as low pass filter for temperature fluctuations with a measured time constant of $\tau_\mathrm{th}\approx8$~d.
%Additional low pass filtering of remaining fluctuations is achieved by two nested passive heat shields made from polished aluminum that surround the cavity %with a measured time constant of approximately $\tau_\mathrm{therm}=11$~d.
%
\begin{figure}[htbp]
\centerline{\includegraphics[width=0.95\columnwidth]{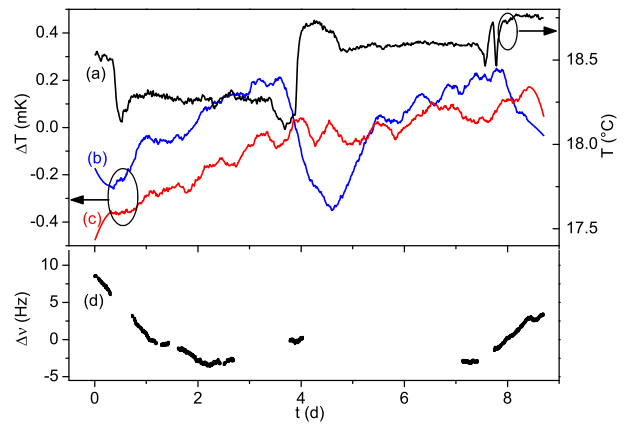}}
\caption{Temperature in the acoustic isolation box surrounding the vacuum chamber (a), temperature fluctuation on the active heat shield (b) and next to the cavity (c). 
(d) shows residual cavity frequency fluctuations measured against the Sr reference with a linear drift of $\SI{+20}{\milli\hertz/\second}$ subtracted.}
\label{fig:temperature}
\end{figure}
With this advanced temperature control, the temperature fluctuations of the outer passive shield remain within  $\approx 500~\mu$K when the room temperature changes by $0.5$~K (Fig.~\ref{fig:temperature}). 
At the resonator no significant influence was observed.
After removal of an average drift of $\SI{+20}{\milli\hertz/\second}$ due to aging of the spacer material, the frequency of the resonator fluctuates by $\SI{\pm6}{\hertz}$.
Assuming an offset of the resonator temperature from %the zero-crossing temperature 
$T_0$ of 50~mK, these frequency deviations correspond to temperature fluctuations of $\pm\si{120}~\mu$K.

%\section{Optical setup}
This resonator setup is the central part of the interrogation laser system for a $^{87}$Sr optical lattice clock \cite{fal14}.
The frequency of a filter-stabilized $\SI{698}{\nano\meter}$ extended-cavity laser diode \cite{bai06,gil07} is locked via the Pound-Drever-Hall (PDH) technique to the cavity \cite{dre83}. 
The light is sent through an offset-AOM, which allows tuning the frequency to the $^{87}$Sr clock transition, and via a $\SI{2}{\meter}$ long optical fiber to the vibration isolation table.
There, the light is phase modulated by a free space electro-optic modulator (EOM) required for PDH-stabilization and then sent through several windows to the resonator.   
Any reflection from optical components in the optical path back to the cavity leads to frequency pulling of the cavity resonance. By placing a quarter wave plate inside the vacuum system directly in front of the cavity mirror, all these reflections have orthogonal polarization with respect to the initial polarization and should not contribute to frequency pulling.

Faraday isolators, $\SI{35}{\decibel}$ before and $\SI{38}{\decibel}$ behind the EOM suppress effects of  spurious reflections, which may also produce residual amplitude modulation (RAM). 
The potential frequency shift due to RAM is on the order of $10^{-17}$, thus no active control was implemented. 
Significant frequency shifts are introduced by thermal deformation and optical length change caused by light absorbed in the mirror coatings. 
It leads to a sensitivity of $\SI{120}{\hertz}$ per $\mathrm{\mu W}$ of power transmitted through the cavity. 
We send a power of $35~\mu$W to the resonator about $70\%$ of the carrier is coupled in and $2~\mu$W are transmitted. 
The transmitted power was stabilized with the offset-AOM. 
An out-of-loop measurement of the power stability converts into a frequency instability below $5\times10^{-17}$ at $\SI{1}{\second}$ averaging time. 
The calculated instability from the shot noise for both the transmitted as well as the light for the PDH lock is well below $10^{-17}$.  

The optical path lengths (including optical fibers) to the reference cavity, to the Sr experiment, and to a frequency comb are all stabilized, using a common reference mirror near the clock laser.
Special attention is paid to the optical path between the reference mirror and the cavity. 
Since it is inside the PDH-stabilization loop, any phase noise $\phi_\mathrm{path}$ from length fluctuations is added to the laser light at the reference mirror and to the light sent elsewhere.
%
%Through the PDH lock, the light at the cavity has the smallest phase noise.
%Any phase noise $\phi_\mathrm{path}$ picked up on the fiber will be interpreted as laser noise and servoed on the laser light.
%
A path length stabilization by retro-reflecting light from the cavity back through the fiber \cite{ma94} is incompatible with the required optical isolation as described above. 
To circumvent this dilemma, we deliver phase-stable light with an additional noise canceled fiber using the common reference mirror and superimpose it with the light transmitted through the reference cavity. 
Phase fluctuations in the heterodyne beat signal originate from $\phi_\mathrm{path}$ and are compensated through the offset-AOM.
%
%To circumvent this dilemma, we deliver phase stable light to the transmitting side of the cavity through an additional noise canceled fiber, which is using the common reference mirror. 
%This light is frequency detuned from the cavity resonance and superimposed with the transmitted light to form a heterodyne beat signal. Detected phase fluctuations originate from $\phi_\mathrm{path}$ and are canceled through a phase locked loop (PLL) steering the offset-AOM.      
%
%Any phase fluctuations through the fiber and free-space paths to the cavity $\phi_\mathrm{path}$ are added to the laser phase noise $\phi_\mathrm{l}$. 
%A common fiber phase noise canceling by retro-reflecting light back through the fiber is strongly not recommended \cite{ma94}. Circumventing this problem, we have integrated an additional common noise canceled fiber, delivering phase stable light to the transmitted side of the cavity from teeing of some light directly from the laser diode. This light is superimposed with the transmitted light to form a heterodyne beat signal on a fast photo diode. Detected phase fluctuations slower than the cavity ring-down time, arises through the fiber noise $\phi_{fib}$. A correction frequency signal is derived and also applied to the offset-AOM.         
%%
%\section{Results}
%
\begin{figure}[htbp]
\centerline{\includegraphics[width=0.95\columnwidth]{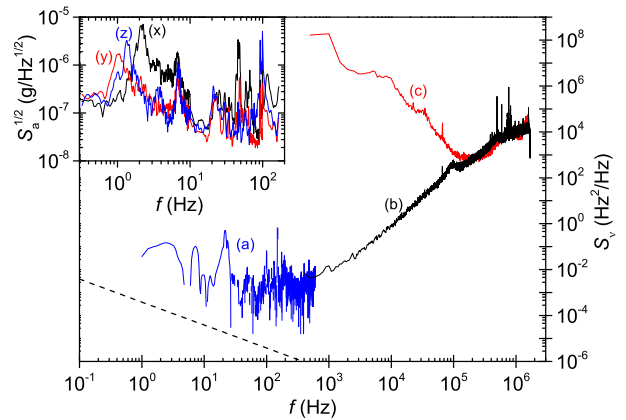}}
\caption{Noise spectrum of the laser locked to the cavity from a three cornered hat method in the frequency domain (a), 
from the light transmitted through the cavity (b), of the free running laser (c) and 
calculated cavity thermal noise (dashed black line). 
For details see text.
The inset displays the power spectral density of acceleration $S_{\textit{a}}$ on the passive vibration isolation table close to the cavity, in $x,y,z$-direction.}
\label{fig:frequencynoise}
\end{figure}

The system performance was evaluated by comparison to a laser locked to a cryogenic silicon cavity at $\SI{1542}{\nano\meter}$ \cite{kes12a} via a frequency comb and a $\SI{698}{\nano\meter}$ transportable laser system based on a $\SI{12}{\centi\meter}$ long ULE-cavity with fused silica mirrors. 
From the simultaneously sampled two optical beats between these systems, the spectral power densities of frequency fluctuations $S_\textit{y}(f)$ between all three systems were calculated.
As these statistical spectral quantities behave the same way as Allan variances $\sigma^2_\textit{y}$, 
the individual noise components $S_\nu(f)$ were computed with the three-cornered hat method \cite{gra74} (Fig.~\ref{fig:frequencynoise}a).
For higher Fourier frequencies, the noise was estimated from the beat between the spectrally filtered light transmitted through the cavity with light coming directly from the laser (b). At high frequencies, this noise approaches the frequency noise of the free running laser (c).

The inset in Fig.~\ref{fig:frequencynoise} shows the acceleration spectrum perturbing the cavity. 
Frequency features at $\SI{1}{\hertz}$ and $\SI{7}{\hertz}$ from resonances of the passive table can also be identified in the laser spectrum. 
The laser noise peak at $\SI{20}{\hertz}$, which is not clearly visible in the vibration spectra, presumably originates from a mechanical resonance of the resonator mounted on the Teflon posts. 

\begin{figure}[htbp]
\centerline{\includegraphics[width=0.95\columnwidth]{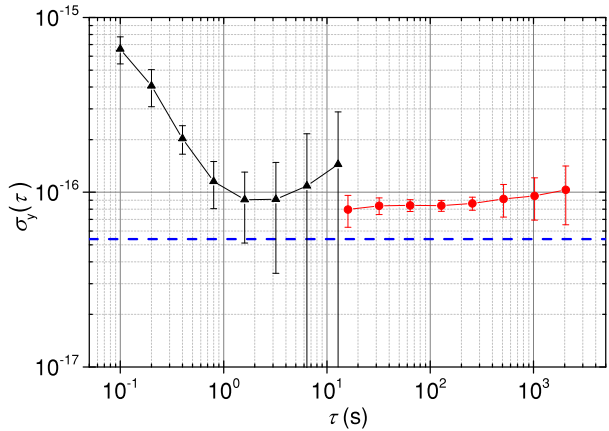}}
\caption{Allan deviation (overlapping) of the cavity-stabilized laser system from three-cornered hat method (black triangles), and from comparison to the Sr reference (red circles) with linear drift removed.
The dashed line indicates the calculated thermal noise level.}
\label{fig:stability}
\end{figure}

The stability of the laser system for averaging times between $\SI{100}{\milli\second}$ to $\SI{10}{\second}$ was calculated with the three-cornered hat method with the two lasers mentioned above (Fig.~\ref{fig:stability}). 
Frequency data were taken by three frequency counters over $\SI{10}{\hour}$ with a gate time of $\SI{100}{\milli\second}$ and a single linear drift was subtracted. 
The data set was cut in over $100$ sections and the individual Allan deviations were calculated.
 
The stability from $10$~s to $2000$~s was calculated by analyzing the offset frequency from the Sr transition shown in Fig.~\ref{fig:temperature}d. 
The data set of about $240\:000$~s was cut in $10\:000$~s long sections, a linear drift for each section was removed for compensating the temperature drift, and the Allan deviations were calculated.
For both methods, the arithmetic mean of the deviations are shown in Fig.~\ref{fig:stability}.
The measured instability of $8\times 10^{-17}$ nearly reaches the calculated thermal noise floor of $5.4\times 10^{-17}$. 
The outstanding frequency stability of below $10^{-16}$ for averaging times from $\SI{1}{\second}$ up to $\SI{1000}{\second}$ indicates high seismic noise suppression and good thermal control.
We assume that the stability at short averaging times is still limited by seismic noise and the free running laser linewidth in combination with the PDH servo bandwidth of $\SI{0.8}{\mega\hertz}$. 

%\section{Conclusion and Outlook}
We have presented to our knowledge the first frequency stabilized laser setup that shows a stability below $1\times10^{-16}$ over a wide range of averaging times, close to the thermal noise limit of this $48$~cm long cavity. 
Both the observed flicker floor near the thermal limit and the deviations from a predictable linear drift are comparable to cryogenic ($T=124$~K) single crystal silicon resonators \cite{kes12,hag14}. 
This was possible by implementing a novel vibration insensitive mount, temperature control, and optical path length control all the way to the cavity mirror. With laser systems like this, optical clocks will be improved by reducing the Dick effect and allowing Fourier limited linewidths for interrogation times beyond one second.

This work was supported by the Centre of Quantum Engineering and Space-Time Reseach (QUEST), the German Research Foundation (DFG) through the RTG~1729 \lq Fundamentals and Applications of ultra-cold Matter\rq \ and the CRC 1128~geo-Q \lq Relativistic geodesy and gravimetry with quantum sensors\rq,~the European Commission's FP7 within SOC2 and the European Metrology Research Programme (EMRP) under QESOCAS and ITOC. 
The EMRP is jointly funded by the EMRP participating countries within EURAMET and the European Union.

We thank E.~Rasel's group at LUH for providing the laser and E.~Tiemann's group, from LUH, for providing the ULE glass spacer and also numerous people from PTB's mechanical, electronic workshop for help and advice as well as Ch.~Tamm for fruitful discussions.  

%
%\bibliographystyle{osajnl}
%\bibliography{texbi431}
%\end{document}

%\bibliographystyle{osajnl}
%\bibliography{texbi431}

\end{document}